\documentclass[useAMS,usenatbib]{mn2e} \usepackage{psfig}
\usepackage{graphicx} \usepackage{multirow} \usepackage{subfigure}

\defcitealias{hunter94_ocon}{Hunter, O'Connell \& Gallagher (1994a)}
\defcitealias{hunter94_vanwoerd}{Hunter, van Woerden \& Gallagher (1994b)}

\defcitealias{schaerer99a}{Schaerer, Contini \& Kunth (1999a)}
\defcitealias{schaerer99b}{Schaerer, Contini \& Pindao 1999b}

\defcitealias{sirianni05}{Sirianni~et~al.~(2005)}

\title [Cluster and nebular properties of NGC~1140]{Cluster and
nebular properties of the central star-forming region of NGC~1140
\thanks{Based on observations collected at the European Southern
Observatory, Chile, under programme ESO 71.B-0058(A), and on
observations obtained with the NASA/ESA \emph{Hubble Space Telescope},
which is operated by the Association of Universities for Research in
Astronomy, Inc., under NASA contract NAS 5-26555.}}

\author[S. L. Moll et
al.]{S. L. Moll$^{1}$\thanks{E-mail:s.moll@sheffield.ac.uk},
S. Mengel$^{2}$, R. de Grijs$^{1,3}$, L. J. Smith$^{4,5}$ and
P. A. Crowther$^{1}$ \\ 
$^{1}$ Department of Physics and Astronomy, University of Sheffield, 
Sheffield S3 7RH \\ 
$^{2}$ European Southern Observatory, D-85748 Garching, Germany \\ 
$^{3}$ National Astronomical Observatories, Chinese Academy of Sciences, 
Beijing 100012, China \\
$^{4}$ Space Telescope Science Institute and European Space Agency,
Baltimore, MD 21218, U.S.A. \\ 
$^{5}$ Department of Physics \& Astronomy, University College London, 
London WC1E 6BT}

\begin{document}

\date{}


\maketitle


\begin{abstract}
We present new high spatial resolution \emph{HST}/ACS imaging of
 NGC~1140 and high spectral resolution VLT/UVES spectroscopy of its
 central star-forming region.  The central region contains several
 clusters, the two brightest of which are clusters 1 and 6 from
 Hunter, O'Connell \& Gallagher, located within star-forming knots A and
 B, respectively.  Nebular analysis indicates that the knots have an
 LMC-like metallicity of $12~+~\log{\mathrm {O/H}}=8.29\,\pm\,0.09$.
 According to continuum subtracted $\mathrm {H} \alpha$ ACS imaging,
 cluster~1 dominates the nebular emission of the brighter
 knot~A. Conversely, negligible nebular emission in knot~B originates
 from cluster~6.  Evolutionary synthesis modelling implies an age of
 $5\,\pm\,1$ Myr for cluster~1, from which a photometric mass of
 $(1.1\,\pm\,0.3) \times 10^6\,\mathrm{M}_\odot$ is obtained. For this
 age and photometric mass, the modelling predicts the presence of
 $\sim 5900$ late O stars within cluster~1.  Wolf-Rayet features are
 observed in knot~A, suggesting 550 late-type WN and 200 early-type WC
 stars. Therefore, $N(\mathrm{WR})/N(\mathrm{O}) \sim 0.1$, assuming
 that all the WR stars are located within cluster~1.  The velocity
 dispersions of the clusters were measured from constituent red
 supergiants as $\sigma \sim 23\,\pm\,1\,\mathrm{km\,s}^{-1}$ for
 cluster~1 and $\sigma \sim 26\,\pm\,1\,\mathrm{km\,s}^{-1}$ for
 cluster~6. Combining $\sigma$ with half-light radii of $8\,\pm\,2$ pc
 and $6.0\,\pm\,0.2$ pc measured from the F625W ACS image implies
 virial masses of $(10\,\pm\,3)\times 10^6\,\mathrm{M}_\odot$ and
 $(9.1\,\pm\,0.8)\,\times\,10^6\,\mathrm{M}_\odot$ for clusters 1 and
 6, respectively.  The most likely reason for the difference between
 the dynamical and photometric masses of cluster~1 is that the
 velocity dispersion of knot A is not due solely to cluster 1, as
 assumed, but has an additional component associated with cluster~2.
\end{abstract}


\section{Introduction}

Violent bursts of star formation, which are characteristic of
starburst galaxies, resemble the star-forming phase of young galaxies
in the early Universe. Nearby starbursts provide local templates to
which distant star-forming galaxies may be directly compared. Only a
handful of starburst galaxies are located within 10 Mpc, yet they
produce around a quarter of the entire high-mass star population
\citep{heckman98_proc}. Thus, starbursts are ideal sites in which to
study massive stars.
The signatures of Wolf-Rayet (WR) stars, which are the highly evolved
descendants of massive O stars, are apparent in a subset of starburst
galaxies, known as WR galaxies. Since WR stars are exclusively
associated with young ($\sim 5$ Myr) stellar populations, WR
galaxies represent young or ongoing starbursts.
A hallmark of all starbursts seems to be the production of luminous,
compact star clusters. The sizes, luminosities and masses of these
Young Massive Clusters (YMCs) are consistent with the properties
expected of young globular clusters (GCs). This led to the suggestion
that YMCs represent globular clusters at an early phase of their
evolution. Dynamical mass measurements can potentially be used to test
the scenario with respect to the long-term survivability of these YMCs
(e.g. \citealt{ho96a,ho96b}; see \citealt{degrijs07_parm} for an
overview). Understanding the role of YMCs as candidate proto-GCs is
vital for our understanding of galaxy formation and evolution, as well
as large-scale star formation.
Clusters also have the advantage of being simple to model -- they can
be approximated as a coeval, simple stellar population with a single
metallicity.  This makes them ideal as probes of burst properties such
as age, duration, chemical evolution and star-formation rate as well
as constraining the parameters of the stellar initial mass function
(IMF).

NGC~1140 is a low-metallicity WR galaxy at a distance of $\sim 20$
Mpc\footnote{Based on the heliocentric velocity of the galaxy,
corrected for the Virgocentric flow, and assuming $H_0 = 70 \,\mathrm{
km\,s}^{-1} \mathrm{Mpc}^{-1}$; adopted from the \emph{HyperLeda}
database at http://leda.univ-lyon1.fr/} and is a prime example of a
nearby analogue of the star-forming galaxies identified at high
redshifts. HI and optical observations indicate that the galaxy has
undergone a merger within the past 1 Gyr and it is thought that this
event is responsible for the plethora of YMCs that the galaxy hosts.
Star clusters have been imaged with both the Wide Field/Planetary
Camera (WF/PC) and the Wide Field Planetary Camera 2 (WFPC2) on the
\emph{Hubble Space Telescope (HST)} by \citetalias{hunter94_ocon} 
\defcitealias{hunter94_ocon}{Hunter et al. (1994a)} and 
\citet{degrijs04_1140}. They identified eight young luminous clusters
in the centre of the galaxy, and their studies indicated that the
clusters have masses of up to a few~$\times 10^6\,\mathrm{M}_\odot$
and ages that lie in the range of a few to a few tens of Myr.

 This paper considers the properties of the two brightest clusters
within NGC~1140, clusters 1 and 6, and the two star-forming knots in
which these clusters are contained. It is structured as follows.
Details of our VLT and \emph {HST} observations and data reduction are
given in Section 2, along with photometry of the clusters.  In
Sections 3 and 4, we discuss the stellar and nebular properties of the
two knots apparent in the VLT spectra, and in Section 5 we present the
results of evolutionary synthesis modelling of cluster~1. The massive
star population of cluster~1 and of both knots~A and B, and the
star-formation rate of NGC~1140, are considered in Section 6. The
dynamical masses of the clusters are determined in Section 7 and we
discuss our findings in Section 8. Finally, we summarise our results
in Section 9.


\section{Observations and Data reduction}

\begin{figure*}
\centerline{\psfig{figure=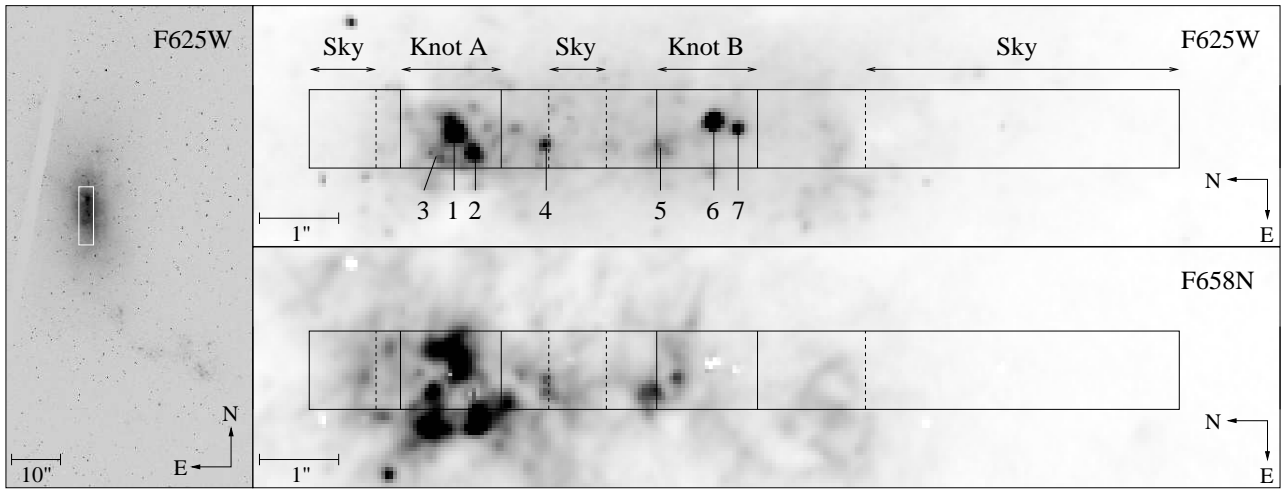,width=17cm,angle=0.}}
\caption{\label{fig:obs} \emph{HST}/ACS images of NGC~1140, at
different cut levels (1 arcsec $\sim 100$ pc), marked as being either
the F625W image or the continuum subtracted F658N image.  The
left-hand image shows the whole galaxy. The white box indicates the
region that is magnified in the images in the right-hand panel. The
lighter diagonal  band across the image is the inter-CCD gap between the
two detectors of ACS.  Superimposed on each right-hand image is the $1
\times 11\,\mathrm{arcsec}^2$ VLT/UVES slit. The regions extracted as
knots~A and B and as sky are indicated by the solid and dashed lines, respectively.
 Clusters $1-7$ \defcitealias{hunter94_ocon}{Hunter et al. 1994a} \protect
\citepalias{hunter94_ocon} \defcitealias{hunter94_ocon}{Hunter et
al. (1994a)} are also labelled. The top right-hand image shows that
clusters 3 and 5 likely comprise several smaller clusters.}
\end{figure*}

We obtained high-resolution spectroscopy of the central region of
 NGC~1140 with the VLT/Ultraviolet and Visual Echelle Spectrograph
 (UVES), in addition to high spatial resolution \emph {Hubble Space
 Telescope (HST)} / Advanced Camera for Surveys (ACS) imaging of
 NGC~1140. Fig. \ref{fig:obs} shows the region of the galaxy observed
 with UVES and labels clusters $1-7$, as designated by
 \citetalias{hunter94_ocon}. It shows the location of the $1 \times
 11\,\mathrm{arcsec}^2$ UVES slit, which was aligned north-south over
 the two knots in the central region of NGC~1140, hereafter called
 knots A and B. Knot~A, which contains the clusters 1 and 2, is the
 brightest region in the optical and near infrared.  Knot~B lies $\sim
 3$ arcsec south of knot~A, and hosts clusters 5, 6 and 7.

\subsection{Imaging of NGC~1140}
\label{sec:imaging}

 NGC~1140 was observed on 21 August 2003 with ACS/WFC aboard \emph
 {HST} as part of GO programme 9892 (PI Jansen).  Two exposures were
 taken -- one of 950s duration with the narrow-band F658N filter and
 one of 100s duration using the broad-band F625W filter. The data were
 reduced using the on-the-fly reduction pipeline.

At the distance of NGC~1140, the F658N filter does not include the
[N~II] 6583 line, and, therefore, is essentially an $\mathrm{H} \alpha$
plus continuum filter. We have  corrected for the underlying
continuum by subtracting the scaled F625W image such that a background
galaxy is removed. This continuum subtracted F658N is included in
Fig. \ref{fig:obs}.  The high resolution of the ACS data also shows
that clusters 3 and 5 from \citetalias{hunter94_ocon} are not single
clusters.

Aperture photometry was carried out on the drizzled F625W ACS image in
 Starlink's {\sc gaia}. Since the crowded nature of the field makes it
 difficult to position apertures to include all of the light from the
 desired cluster without contamination from its neighbours, several
 circular apertures with a range of radii and positions were
 considered for each region.  Each measurement was sky subtracted
 using a mean sky value obtained from several small apertures of sky,
 and aperture corrected using the values contained in
 \citetalias{sirianni05}. As these aperture corrections are valid for point
 sources rather than for spatially extended clusters, our photometry
 may be an underestimate. However, since the light profile of the
 clusters is difficult to ascertain (see Section \ref{sec:rhl}), no
 correction for this effect is made.

Table~\ref{tab:mag} presents our F625W photometry.  The associated
uncertainty represents the range of results produced.  Table
\ref{tab:mag} also contains F300W and F814W photometry.  The F300W and
F814W photometry of knots A and B were determined using archival WFPC2
images (GO 8645, PI Windhorst) using the same method as described for
the F625W images. The F300W and F814W cluster photometry was taken
from \citet{degrijs04_1140}. The pre-COSTAR WF/PC photometry of
\citet{degrijs04_1140} and \citetalias{hunter94_ocon} was not included
due to its lower spatial resolution.

\begin{table}
\caption{\label{tab:mag} F625W ACS aperture {\sc stmag} photometry of
the brightest central clusters of NGC~1140 and the regions designated
knots A and B. The F300W and F814W cluster magnitudes are taken from
WFPC2 imaging of \protect \citet{degrijs04_1140}.}
\begin{tabular}{|l|c|c|c|} \hline
Cluster & $m_{\mathrm{F300W}}$ & $m_{\mathrm{F625W}}$ &
                $m_{\mathrm{F814W}}$ \\ & (mag) & (mag) & (mag) \\
                \hline Cluster~1 & $16.50\,\pm\,0.05$ &
                $17.9\,\pm\,0.1$ & $18.51\,\pm\,0.05$ \\ Cluster~2 &
                $16.37\,\pm\,0.05$ & $18.8\,\pm\,0.1$ &
                $19.11\,\pm\,0.08$ \\ Cluster~6 & $18.47\,\pm\,0.09$ &
                $18.4\,\pm\,0.1$ & $18.73\,\pm\,0.03$ \\ Cluster~7 &
                $21.71\,\pm\,1.51$ & $19.7\,\pm\,0.1$ &
                $19.86\,\pm\,0.04$ \\ Knot~A & $15.5\,\pm\,0.1$ &
                $17.0\,\pm\,0.1$ & $17.8\,\pm\,0.1$ \\ Knot~B &
                $17.6\,\pm\,0.1$ & $17.5\,\pm\,0.1$ & $18.4\,\pm\,0.1$
                \\ \hline
\end{tabular}  
\end{table}

\subsection{Spectroscopy of NGC~1140}

The central region of NGC~1140 was observed with UVES on the VLT
Kueyen Telescope (UT2) in Chile on 19-20 September 2003.  UVES is a
cross-dispersed echelle spectrograph, with two arms -- a blue arm
comprising a single EEV CCD and a red arm comprising a mosaic of an
EEV and a MIT-LL CCD.  Using the dichroic beam splitter, red and blue
data were taken simultaneously: cross-disperser $\#4$ with
$312\,l\,\mathrm{mm}^{-1}$ was centred on 840 nm and cross-disperser
$\#2$ with $660\,l\,\mathrm{mm}^{-1}$ was centred on 437 nm.  Thus,
data from the regions $\sim 3740-4985\textrm{\AA}$ and $\sim
6475-10090\textrm{\AA}$ were obtained.  The spectral region $\sim
8180-8420\textrm{\AA}$ was not observed since it lay on the gap
between the two red CCDs. The data were $2 \times 2$ binned, so that
the red data have a pixel scale of 0.364 arcsec pixel$^{-1}$ and the
blue data have a scale of 0.512 arcsec pixel$^{-1}$.

The galaxy was observed with UVES for a total time of 14700s on the
 first night and 16200s on the second night.  The seeing averaged
 $\sim 0.8$ arcsec on the first night, varying by $\sim 0.1$ arcsec
 and was somewhat poorer on the second night.  The two knots were more
 clearly resolved in the data obtained on 19 September and so only the
 data of NGC~1140 from this night were considered in the analysis.

A telluric star, HIP 11918, was also observed in the same setup. Flux
  standards were not part of the original observing
  programme. However, we obtained data of two flux standards, EG274
  and EG21. EG274 was observed for another programme on the first half
  of the first night for 600s using the blue setup above, plus a red
  setup centred on $8600 \textrm{\AA}$. EG21 was observed on the
  second night in 4 exposures of 2s in only the blue setup.  ThAr arcs
  were taken for wavelength calibration.  The resolution of the data,
  as measured from the ThAr arc lines, is $\sim
  7\,\mathrm{km\,s}^{-1}$.

The data were reduced using the software package {\sc iraf}.  The data
 were bias corrected with a median bias frame, and divided by a
 normalised flat field. The two knots were extracted, each with an
 aperture of $\sim 1.3$ arcsec, and background subtracted using a
 polynomial fit to the sky regions indicated in
 Fig. \ref{fig:obs}. After extraction, median knot spectra were
 created and wavelength calibrated, before the orders were merged.
 The standard stars were similarly extracted.  This method
 over-subtracts the sky in the region of nebular lines. Therefore, the
 NGC~1140 spectra were also extracted without sky subtraction. The sky
 level was then manually subtracted with a polynomial fit so that the
 continuum matched that of the sky-subtracted spectra away from the
 nebular lines.

\begin{figure}
\includegraphics[angle=0,width=8.5cm]{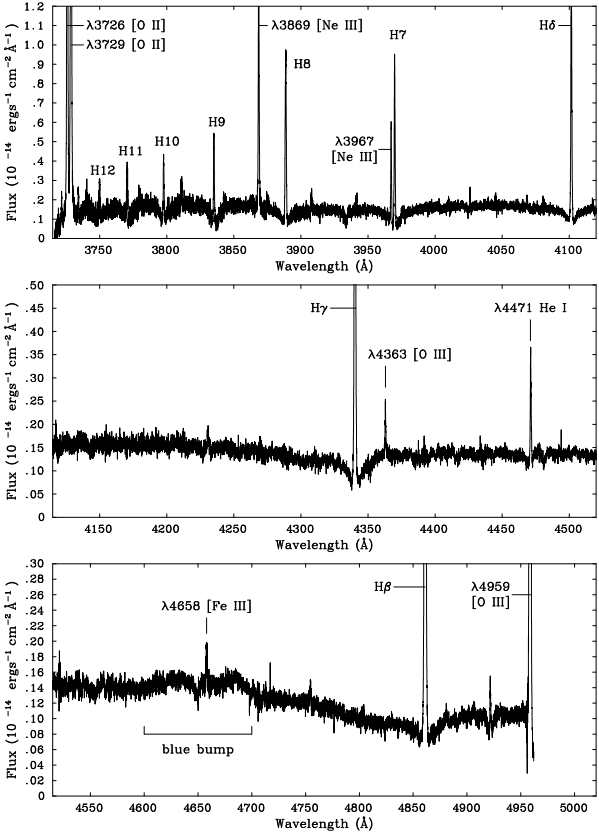}
\caption{\label{fig:spectra} Blue spectrum of knot~A with the
principal nebular lines and the Wolf-Rayet blue bump marked, corrected
to the stellar rest-frame.  Broad stellar absorption is apparent
underlying the narrow nebular Balmer emission.}
\end{figure}

The blue NGC~1140 data were flux calibrated with EG274, and adjusted
 to match the F625W photometry of knot~A and B contained in Table
 \ref{tab:mag}.  Due to the lack of a suitable red flux standard, it
 was not possible to flux calibrate the red data.  The calibrated blue
 spectrum of knot~A is shown in Fig. \ref{fig:spectra}, with the
 principal nebular lines indicated.

\subsection{Spectroscopy of red supergiants}

Comparison red supergiant (RSG) spectra are required to determine a
cluster's velocity dispersion based upon its RSG features, which can
then be used to calculate its dynamical mass. Therefore, six Small
Magellanic Cloud (SMC) red supergiants, taken from the
\citet{massey03} catalogue, were observed with the setup described
above.  The catalogue ID number of these stars, and their spectral
types were 20133 (M0 I), 64448 (K2-7 I), 71566 (K7 I), 50840 (M1-2 I),
66694 (K5 I) and 71507 (K3-5 I).

 
\section{Stellar properties}

Red supergiant and OB stellar features are visible in the UVES spectra
of both knots. Wolf-Rayet (WR) features are also present in the
spectrum of knot~A.

\subsection{OB and WR star features}

Knots A and B both show stellar absorption in the hydrogen Balmer
series, arising from the presence of early-type stars, underlying
strong nebular emission.  The equivalent widths of the underlying
stellar absorption are $\mathrm{W}_\lambda (\mathrm{H} \beta) \sim 4
\textrm{\AA}$, $\mathrm{W}_\lambda (\mathrm{H} \gamma) \sim 3
\textrm{\AA}$ and $\mathrm{W}_\lambda (\mathrm{H} \delta) \sim 3
\textrm{\AA}$.

 WR stars are the bare cores of O stars, in their final evolutionary
 stages, characterised by strong, broad emission lines. Their strong
 winds reveal first the CNO-burning products, observable as
 nitrogen-rich (WN) stars and subsequently the He burning products,
 producing carbon-rich (WC) stars. A WR galaxy, or a cluster
 containing WR stars, can be identified by the presence of the blue
 bump around $\lambda 4686 \textrm{\AA}$ or the yellow bump at
 $\lambda 5808 \textrm{\AA}$ \citepalias{schaerer99b}. The yellow bump
 is produced solely by WC stars, while the blue bump can contain
 features of WN stars, such as stellar N~III 4640 and He~II 4686,
 and/or WC stars, such as C~III 4650 and C~IV 4658, with narrow
 nebular features superimposed, the strongest of which is [Fe~III]
 4658.

 The blue WR bump emission feature in the spectrum of knot~A has an
equivalent width of $\mathrm{W}_\lambda \sim 12.9 \textrm{\AA}$, with
He~II 4686 contributing around $\sim 4.6 \textrm{\AA}$.  The presence
of WR features in the spectrum of knot~A constrains the age of the
knot to $4\,\pm\,1$~Myr \citep{crowther07_rev}.  No WR emission
features are visible in the spectrum of knot~B. The absence of WR
stars implies that knot~B is either younger than around 3 Myr, or
older than around 5 Myr. However, the presence of RSG features in the
spectrum (see Section \ref{sec:RSG}) makes it unlikely that the knot
is younger than 3 Myr, for the case of an instantaneous burst.

\subsection{RSG features}
\label{sec:RSG}

\begin{figure}
\includegraphics[angle=0,width=9cm]{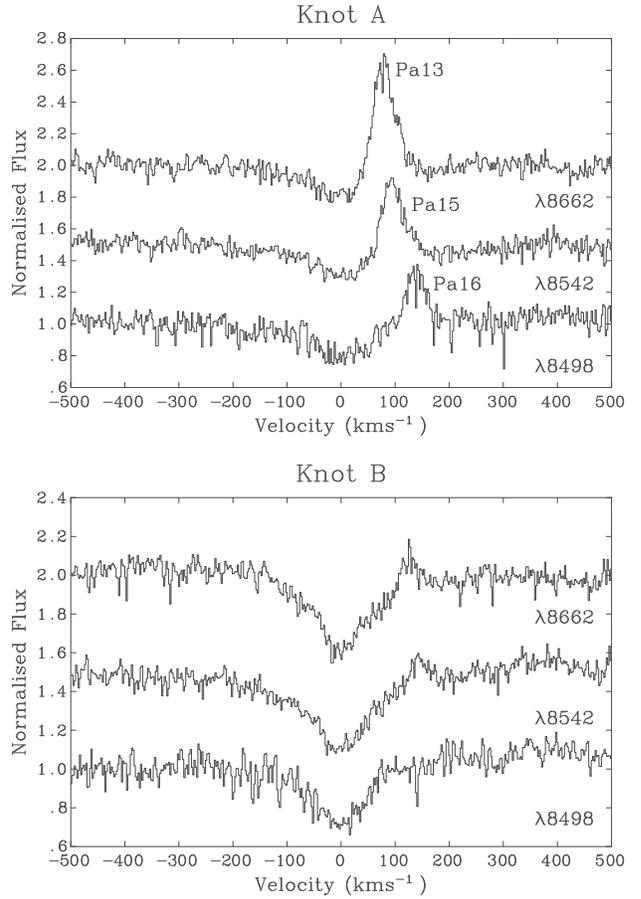}
\caption{\label{fig:CaT} The normalised Ca~II triplet lines of knot~A
(top) and knot~B (bottom) after correction for a recession velocity of
$v_r = 1475\,\mathrm{km\,s}^{-1}$. The $\lambda 8542$ and the $\lambda
862$ lines are offset by 0.5 and 1.0, respectively. Nebular Paschen
emission lines are indicated.}
\end{figure}

The Ca~II triplet absorption lines ($\lambda \lambda 8498, 8542,
8662\,\textrm{\AA}$) are clearly visible in the red spectra of both
knots A and B. These features, along with other weak metal lines
present in the data, such as Fe~I and Mg~I, arise from the RSGs in the
cluster. The presence of RSGs indicates a cluster age greater than
around 5 Myr.

The Ca~II triplet lines of both knots were fitted with Gaussian
profiles with the {\sc elf} (emission-line fitting) routine in
Starlink's {\sc dipso} package. The line centres, widths and intensity
of the fits were allowed to vary freely, and the measured line centres
used to determine the recession velocity of the galaxy.
 A good fit could not be obtained to the $\lambda 8542$ and $\lambda
 8662$ lines of knot~A, due to the presence of Paschen nebular lines
 within these profiles.  Therefore, the recession velocities
 determined from the remaining four lines were averaged to obtain $v_r
 \sim 1475\,\pm 4\,\mathrm{km\,s}^{-1}$.  Our result is in good
 agreement with other recent values of recessional velocity, such as
 $v_r = 1498\,\pm\,33\,\mathrm{km\,s}^{-1}$ from optical measurements
 \citep{devauc91}, $v_r = 1501\,\pm\,1\,\mathrm{km\,s}^{-1}$ from
 21-cm HI observations \citep{haynes98_hogg} and $v_r =
 1480\,\pm\,9\,\mathrm{km\,s}^{-1}$ from [Fe~II] measurements
 \citep{degrijs04_1140}.

After velocity correction, the equivalent widths of the Ca~II triplet
 lines were measured with {\sc elf}, fixing the central wavelengths of
 the lines. This yields values of $\mathrm{W}_\lambda(8498) =
 0.77\,\pm\,0.07 \textrm{\AA}$, $\mathrm{W}_\lambda(8542) =
 1.05\,\pm\,0.21 \textrm{\AA}$ and $\mathrm{W}_\lambda(8662) =
 0.91\,\pm\,0.18 \textrm{\AA}$ for knot~A and
 $\mathrm{W}_\lambda(8498) = 0.92\,\pm\,0.09 \textrm{\AA}$,
 $\mathrm{W}_\lambda(8542) = 2.2\,\pm\,0.3 \textrm{\AA}$ and
 $\mathrm{W}_\lambda(8662) = 1.45\,\pm\,0.15 \textrm{\AA}$ for knot~B.
 The velocity corrected Ca~II triplet lines for knots A and B are
 presented in Fig. \ref{fig:CaT}.

\section{Nebular Properties}

In this section we derive information on the gas dynamics of the knots
from the profiles of the nebular lines. Extinctions, electron
densities, temperatures and elemental abundances are determined from
the relative fluxes of the nebular lines.

\subsection{Line profiles and dynamics}
\label{sec:profile}

\begin{figure}
\includegraphics[angle=-90,width=8.5cm]{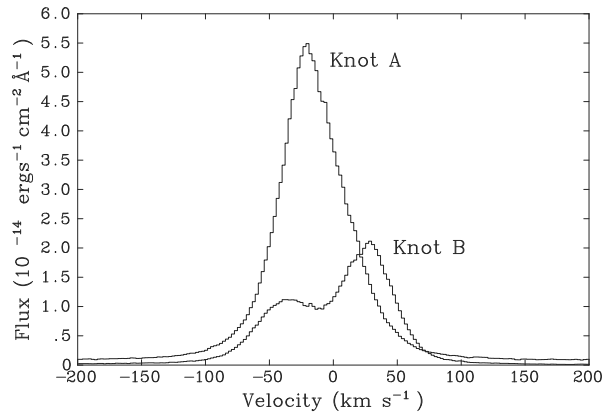}
\caption{\label{fig:hbeta} The H$\beta$ profiles for knots A and B,
after correction to $v_r = 1475\,\mathrm{km\,s}^{-1}$.}
\end{figure}

 Fig. \ref{fig:hbeta} shows that the nebular $\mathrm{H} \beta$
 profiles of knot~A and knot~B are very different.
 Knot~A is dominated by one component at approximately $v \approx
 -15\,\mathrm{km\,s}^{-1}$, but also has a weaker underlying broad
 component. Knot~B more clearly comprises two components: a weaker
 component blueshifted at $v \approx - 36\,\mathrm{km\,s}^{-1}$ with
 FWHM $\sim 58\,\mathrm{km\,s}^{-1}$ and a brighter redshifted
 component at $v \approx 28\,\mathrm{km\,s}^{-1}$ with FWHM $\sim
 47\,\mathrm{km\,s}^{-1}$. These nebular profiles are representative
 of all the strong nebular lines of the knots. Examination of the 2D
 image indicates that there is diffuse emission throughout the central
 region of the galaxy that probably corresponds to the bluer, weaker
 component of knot~B and to the weak, broad component of knot~A. The
 stronger components of the knots arise from more discrete emission
 from the clusters within these knots, as are visible in
 Fig. \ref{fig:obs}. Alternatively, the broad underlying component of
 the knot~A profiles may be due to hot cluster winds impacting on the
 surrounding interstellar clouds, as has been seen in other young
 star-forming regions (see \citealt{sidoli06} for a review and
 \citealt{westmoquette07} for a detailed discussion). Turbulent
 broadening of the order of $30\,\mathrm{km\,s}^{-1}$ was implied from
 the FWHMs of the main components of the strong nebular lines of knots
 A and B.

\subsection{Line fluxes and extinction}
\label{sec:fluxes}

Nebular line fluxes were measured by fitting the nebular emission with
Gaussian profiles using the {\sc elf} routine, allowing line centres,
widths and intensity to vary freely. The strong nebular emission of
both knots A and B was fitted with two Gaussians (Section
\ref{sec:profile}).
\begin{table}
\caption{\label{tab:extinc} Comparison between predicted intrinsic
 flux ratios, $\mathrm{I}_\lambda/\mathrm{I(H}\beta)$ and observed
 flux ratios, $\mathrm{F}_\lambda/\mathrm{F(H}\beta)$.  For the fluxes
 quoted here, both components of the knot~A profile were summed, while
 only the strong redshifted component of knot~B was considered.  The
 observed ratios were first dereddened with a standard Galactic
 extinction law. Internal extinctions, $E(B-V)_{\mathrm{int}}$, were
 then derived from predicted intensity ratios assuming a \protect
 \citet{howarth83} LMC extinction law. }
\begin{tabular}{llccc} \hline
       &   Line              & $\mathrm{F}_\lambda/\mathrm{F(H}\beta)$ & $\mathrm{I}_\lambda/\mathrm{I(H}\beta)$ & $E(B-V)_{\mathrm{int}}$\\
       &                     & (Observed)           & (Predicted)   & (mag)               \\ \hline
Knot~A & $\mathrm{H}\gamma$  & $0.427\,\pm\,0.021$  & 0.468         & $0.16\,\pm\,0.10$   \\
       & $\mathrm{H}\delta$  & $0.225\,\pm\,0.011$  & 0.259         & $0.16\,\pm\,0.07$   \\ \hline
Knot~B & $\mathrm{H}\gamma$  & $0.427\,\pm\,0.021$  & 0.468         & $0.19\,\pm\,0.10$   \\
       & $\mathrm{H}\delta$  & $0.212\,\pm\,0.011$  & 0.259         & $0.27\,\pm\,0.07$   \\ \hline 
\end{tabular}
\end{table}

The internal extinctions of the knots, $E(B-V)_{\mathrm{int}}$, were
determined by comparing the observed flux ratios of
$\mathrm{F}(\mathrm{H}\gamma)/\mathrm{F}(\mathrm{H}\beta)$ and
$\mathrm{F}(\mathrm{H}\delta)/\mathrm{F}(\mathrm{H}\beta)$ to the
intrinsic intensity ratios predicted by assuming Case B recombination
theory for electron densities of $10^2\,\mathrm{cm}^{-3}$ and a
temperature of $10^4\,\mathrm{K}$ \citep{hummer87_storey}.
 As the diffuse emission contained within knot~B is expected to have a
 very low extinction compared to the discrete emission from the
 clusters, the nebular extinction was calculated based only on the
 stronger, redder component of knot~B. Since the division between
 components is less well defined for knot~A, both components were
 summed.
The observed flux ratios were first corrected for foreground
extinction of $E(B-V)_{\mathrm{fore}}=0.038$ mag \citep*{schlegel98}
using a standard Galactic extinction law before determining the
internal extinctions of the knots using a  Large
Magellanic Cloud (LMC) extinction law \citep{howarth83}. 
The ratios measured and
internal extinction calculated are presented in Table
\ref{tab:extinc}.  Ratios of $\mathrm{H} \gamma$ and $\mathrm{H}
\delta$ to $\mathrm{H} \beta$ were considered since $\mathrm{H}\alpha$
was not flux calibrated. Mean values of $E(B-V)_{\mathrm{int}} =
0.16\,\pm\,0.06$ mag for knot~A and $E(B-V)_{\mathrm{int}} =
0.23\,\pm\,0.08$ mag for knot~B were subsequently adopted.

 Table~\ref{tab:flux} presents observed and extinction corrected line
fluxes, normalised to $\mathrm{H}\beta = 100$.  Here, both components
were summed for both knots, as the total emission within the knot was
of interest.  Lines with total fluxes of $< 0.2 \%$ of
$\mathrm{H}\beta$ were not included.  We concentrate on deriving the
electron density, temperature and nebular abundances for knot~A
because of the added complications introduced by the nebular structure
present in Knot~B.
\begin{table}
\caption{\label{tab:flux} Observed ($\mathrm{F}_\lambda$) and
intrinsic ($\mathrm{I}_\lambda$) nebular line fluxes relative to
$\mathrm{H}\beta=100$ for knot~A and knot~B.  The fluxes quoted here
are the summation of both components of the nebular lines, for both
knots A and B.  The intrinsic values are corrected for a foreground
extinction of $E(B-V)_{\mathrm{fore}}=0.038$ mag with a standard
Galactic extinction law and then for internal extinctions of
$E(B-V)_{\mathrm{int}}=0.16\,\pm\,0.06$ mag for knot~A and
$E(B-V)_{\mathrm{int}}=0.23\,\pm\,0.08$~mag for knot~B using the
\citet{howarth83} LMC extinction law. The flux of $\mathrm{H}\beta$ is
in units of $\times
10^{-14}\,\mathrm{erg\,s}^{-1}\,\mathrm{cm}^{-2}$.}
\begin{tabular}{@{\hspace{0mm}}l@{\hspace{2.5mm}}r@{$\pm$}lr@{$\pm$}lr@{$\pm$}lr@{$\pm$}l@{\hspace{0mm}}} \hline
                     &\multicolumn{4}{c}{Knot~A} &  \multicolumn{4}{c}{Knot~B} \\ 
Line                 &\multicolumn{2}{c}{$F_{\lambda}$} &\multicolumn{2}{c}{$I_{\lambda}$} 
     &\multicolumn{2}{c}{$F_{\lambda}$} &\multicolumn{2}{c}{$I_{\lambda}$} \\ \hline
$\lambda$3726 [O~II]     & 65&3      & 82&7      & 81&4      & 112&12 \\ 
$\lambda$3729 [O~II]     & 89&4      & 111&10    & 110&6     & 151&17 \\ 
$\lambda$3869 [Ne~III]   & 21.0&1.0  & 25.5&2.0  & 20.0&1.0  & 26.1&2.5 \\ 
$\lambda$3967 [Ne~III]   & 5.7&0.6   & 6.8&0.8   & 4.8&0.5   & 6.1&0.8 \\ 
$\lambda$4026 He~I       & 0.89&0.18 & 1.05&0.22 & 0.89&0.18 & 1.11&0.23 \\
$\lambda$4069 [S~II]     & 0.95&0.19 & 1.10&0.23 & 2.27&0.23 & 2.8 &0.3 \\ 
$\lambda$4102 H$\delta$  & 22.6&1.1  & 26.0&1.7  & 23.8&1.2  & 28.9&2.2 \\
$\lambda$4341 H$\gamma$  & 42.7&2.1  & 46.8&2.7  & 43.1&2.2  & 49 &3 \\
$\lambda$4363 [O~III]    & 1.5&0.15  & 1.63&0.17 &1.03&0.10  & 1.16&0.12 \\
$\lambda$4471 He~I       & 3.1&0.3   & 3.3&0.3   & 3.4&0.3   & 3.7&0.4 \\
$\lambda$4658 [Fe~III]   & 1.08&0.11 & 1.12&0.11 & 2.22&0.22 &2.33&0.23\\ 
$\lambda$4686 He~II      & 0.59&0.12 & 0.61&0.12 & 0.88&0.18 & 0.91&0.18\\
$\lambda$4702 [Fe~III]   & 0.52&0.10 & 0.53&0.11 &\multicolumn{2}{c}{---}
                                                             &\multicolumn{2}{c}{---}\\ 
$\lambda$4861 H$\beta$   &\multicolumn{2}{c}{100} &\multicolumn{2}{c}{100} 
                            &\multicolumn{2}{c}{100} & \multicolumn{2}{c}{100}\\
$\lambda$4959 [O~III]    & 98&5      & 96&5      & 96&5      & 94&5 \\\hline 
H$\beta$                 & 5.4&0.5   & 10&3      & 2.70&0.27 & 7&3 \\ \hline
\end{tabular}
\end{table}

\subsection{Electron densities and temperatures}
\label{sec:temp}

The electron density, $N_e$, and electron temperature, $T_e$, of
knot~A were determined in the five-level atom calculator {\sc temden}
within {\sc iraf} using the diagnostic line ratios of [O~II]
$\mathrm{I(}\lambda 3726)/\mathrm{I(}\lambda 3729)$ and [O~III]
$\mathrm{I(}\lambda 4363)/\mathrm{I(}\lambda 4959)$. This gives values
of $N_e = 60\,\pm\,50\,\mathrm{cm}^{-3}$ (consistent with the low
density limit) and ${T_e} = 9700\,\pm\,300\,\mathrm{K}$.

\subsection{Abundances}

Knot A abundances were calculated from standard [O~II] and [O~III]
diagnostics, plus the values of $N_e$ and $T_e$ determined in Section
\ref{sec:temp}.  These yield values of
$N(\mathrm{O}^+)/N(\mathrm{H}^+) = (8.2\,\pm\,2.0)~\times~10^{-5}$ and
$N(\mathrm{O}^{2+})/N(\mathrm{H}^+) = (1.10\,\pm\,0.20) \times
10^{-4}$.  These imply an abundance for knot A of $12 +
\log\mathrm{O/H}=8.29\,\pm\,0.09$. We adopt this value for
NGC~1140. It lies between the abundance of the Small Magellanic Cloud
(SMC) of $12 + \log\mathrm{O/H}=8.13$ and that of the LMC of $12 +
\log\mathrm{O/H}=8.37$ \citep{russell90}.  It agrees well with other
direct abundance measurements of NGC~1140 of $8.18 \pm 0.06$
\citep{izotov04} and $8.26 \pm 0.07$ (\citealt*{nagao06}; recalculated
from the measurements of \citealt{izotov04}).

As the determined abundance is most similar to the LMC abundance, we
 adopt an LMC-like metallicity with $Z~=~0.008$ for spectral modelling
 purposes.

\section{Photometrically determined cluster properties}
\label{sec:model}

Having established that knot~A has an LMC-like metallicity and an age
of around 5 Myr, we computed v5.1 Starburst99 evolutionary synthesis
models \citep{leitherer99,vazquez05} for an instantaneous burst of
star formation, with a total stellar mass of $10^6\,\mathrm{M}_\odot$
and a metallicity of $Z~=~0.008$ for ages between 1 and 10 Myr in 0.5
Myr intervals. We adopted a \citet{kroupa02} IMF (with slope $\alpha =
2.3$ for $0.5 \le M/\mathrm{M}_\odot \le 100$ and $\alpha = 1.3$ for
$0.1 \le M/\mathrm{M}_\odot < 0.5$) and Padova stellar evolutionary
tracks \citep{fagotto94}, which include careful consideration 
 of the RSG phase.

\subsection{Age}

The Starburst99 model predicts the equivalent widths for $\mathrm{H}
 \alpha$, $\mathrm{H} \beta$ and the Ca~II triplet.  The equivalent
 widths measured for each knot and the ages implied from the
 Starburst99 model are presented in Table~\ref{tab:age}.

\begin{table}
\caption{\label{tab:age} Equivalent widths and knot ages implied from
the Starburst99 model. The equivalent width of Ca~II is defined as
$\mathrm{W}_\lambda(\lambda 8542)+\mathrm{W}_\lambda(\lambda 8662)$
\protect \citep*{diaz89}.}
\begin{tabular}{lr@{$\pm$}lr@{$\pm$}lr@{$\pm$}lr@{$\pm$}l} \hline
           &\multicolumn{4}{c}{Knot~A} &\multicolumn{4}{c}{Knot~B} \\
Line &\multicolumn{2}{c}{$W_\lambda (\textrm{\AA})$}
&\multicolumn{2}{c}{Age (Myr)} &\multicolumn{2}{c}{$W_\lambda
(\textrm{\AA})$} &\multicolumn{2}{c}{Age (Myr)} \\ \hline $\mathrm{H}
\alpha$ & 289&29 & 5.5&0.5 & 142&14 & 6.0&0.5 \\ 
$\mathrm{H} \beta$ & 57 &6 & 5.5&0.5 & 62&6 & 5.0&0.5 \\ Ca~II
&1.96&0.28 & 5.5&0.5 & 3.7&0.3 &\multicolumn{2}{l}{$\ge 6$} \\ \hline
\end{tabular}
\end{table}

 Fig. \ref{fig:obs} and Table~\ref{tab:mag} help to put these
equivalent widths in context by showing the regions in which the
continuum light and the nebular emission originate. In knot~A, both
the stellar continuum and nebular emission are dominated by cluster~1,
indicating that the $\mathrm{H} \alpha$ and $\mathrm{H} \beta$
equivalent widths are meaningful and can be applied to cluster~1.  On
the other hand, the stellar continuum of knot~B is dominated by
cluster~6, while the nebular emission is due solely to cluster~5 and a
second source of emission $\sim 0.4$ arcsec immediately to the
southwest of cluster~5. Therefore, these latter two sources must be
younger than the $5-6$~Myr quoted in Table~\ref{tab:age}. Indeed, the
lack of Wolf-Rayet features in knot~B indicates that they are younger
than $\sim 3$~Myr.  Fig. \ref{fig:obs} shows no nebular emission from
cluster~6 or 7, so these clusters must be older than $\sim 10$~Myr,
unless any gas associated with these clusters has been removed by
multiple supernovae at an even earlier stage.
The equivalent widths of the Ca~II triplet for both knots A and B are
meaningful and can be applied to cluster~1 and 6, respectively, since
these clusters dominate the continuum in the region. However, only a
lower limit to the age of cluster~6 can be determined from the
equivalent width of the Ca~II triplet, because it does not vary
strongly with age after a few Myr. Furthermore, the contribution of
RSGs to cluster evolution is not well understood
(e.g. \citealt{massey03}), and as such RSGs do not provide a reliable
cluster age indicator.

After considering both the age indication of $4\,\pm\,1$~Myr from the
presence of WR features in the spectrum of knot~A and the realistic
age estimates from Table~\ref{tab:age}, we adopt an age of
$5~\,~\pm~\,~1$~Myr for knot~A.
Since no age can be estimated for cluster~6, it is not considered
further in this section.

We find an age differential between knot~A and knot~B, in agreement
 with the results of both \citetalias{hunter94_ocon} and
 \citet{degrijs04_1140}.  Based on their $V - I$ colours,
 \citetalias{hunter94_ocon} estimated the ages of the clusters in the
 region of knot~A as $\sim 3$ Myr and interpreted the redder colours
 of the knot~B clusters as being due to an older age of these
 clusters, as opposed to a higher extinction. They estimated the ages
 of these clusters as $\sim 15$ Myr, since this was the age determined
 by \citet*{oconnell94} for the YMCs in NGC 1569 and NGC 1705 with the
 same colours.  The near infrared CIRPASS data of
 \citet{degrijs04_1140} showed strong [Fe~II] emission throughout the
 galaxy, while strong Br(12-4) and Br(14-4) was predominantly confined
 to knot~A. Since the diffuse [Fe~II] emission is likely associated
 with supernova remnants, while the Brackett nebular emission lines
 are associated with HII regions, \citet{degrijs04_1140} argued that
 the ratio of these lines is a good age indicator. This would suggest
 that knot~B is several Myr older than knot~A.

\subsection{Extinction}
\label{sec:ext}

\begin{figure}
\includegraphics[angle=-90,width=8.5cm]{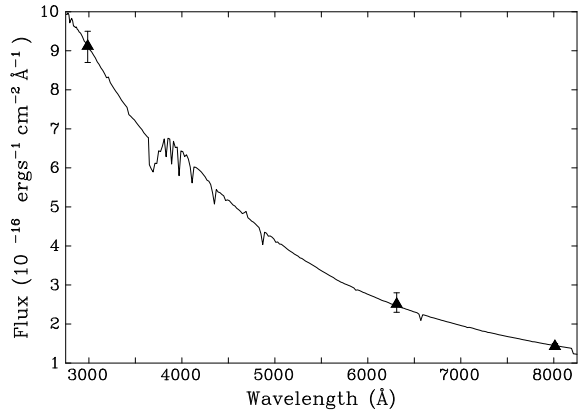}
\caption{\label{fig:SED} Fit of the reddened 5 Myr old model spectrum
(Starburst99) to ACS and WFPC2 photometry for cluster~1. The model
spectrum was first reddened by the foreground extinction of
$E(B-V)_{\mathrm{fore}} = 0.038$~mag \citep{schlegel98} using a
standard Galactic reddening law and then further reddened with a
\citep{howarth83} LMC extinction law to best fit the plotted
photometry.}
\end{figure}

The model spectrum produced by the Starburst99 model for a 5 Myr old
instantaneous burst was reddened to reproduce the ACS and WFPC2
photometry of cluster~1 (Table~\ref{tab:mag}). The best fit, presented
in Fig. \ref{fig:SED}, yields a value of internal cluster reddening of
$E(B-V)_{\mathrm{int}} \sim 0.24\,\pm\,0.04$ mag for cluster~1.  The
uncertainty quoted here solely considers the photometric
uncertainties. The F300W and F625W photometry was more heavily
weighted than the F814W photometry, since these are more sensitive to
reddening. Our value agrees well with the result of
\citet{degrijs04_1140} of $E(B-V)_{\mathrm{int}}= 0.14-0.26$ mag for
cluster~2, determined from spectral energy distribution fits to their
WF/PC and WFPC2 photometry. It also agess with the nebular extinction
determined from the Balmer lines, given in Table~\ref{tab:extinc}.

\subsection{Photometric mass}

The F625W magnitude for cluster~1 was converted into an apparent
$V$-band magnitude using the transformation in \citetalias{sirianni05},
assuming the Starburst99 model $V-R$ colour for a 5 Myr old burst and
adopting the cluster extinction determined in Section \ref{sec:ext}.
A distance of 20 Mpc to NGC~1140 then implies a $V$-band luminosity of
$L_V~=~(7.3\,\pm\,1.8) \times 10^7 L_\odot$ for cluster~1. Comparing
this luminosity to the Starburst99 prediction yields a photometric
mass of $(1.1\,\pm\,0.3) \times 10^6\,\mathrm{M}_\odot$.  This is
based on a Kroupa IMF.

\section{Massive star population and star-formation rate of NGC~1140}

In this section we determine the O star content of the knots from the
$\mathrm{H} \beta$ luminosity of the UVES spectroscopy. We estimate
the WR content of knot~A from the blue bump seen in its spectrum and
measure the star formation rate (SFR) of NGC~1140 from the continuum subtracted F658N ACS
imaging.

\subsection{O star content}

An estimate of the number of O stars within an individual knot can be
obtained from the $\mathrm{H}\beta$ luminosity, $L(\mathrm{H}
\beta)$. Assuming Case B recombination theory \citep{hummer87_storey}
and that a `normal' stellar population is the only source of ionising
photons within the knot, the number of equivalent O7V stars,
$N(\mathrm{O7V})$, contained within the knot is given by:
$$ N(\mathrm{O7V}) = \frac{Q_0^{\mathrm{Obs}}}{Q_0^{{\mathrm{O7V}}}} =
\frac{L(\mathrm{H} \beta) \times 2.1 \times
10^{12}}{Q_0^{\mathrm{O7V}}}
$$
%
(see e.g. \citealt{vacca94b}). Here, $Q_0^{\mathrm{Obs}}$ is the
observed total Lyman continuum luminosity of the knot and
$Q_0^{\mathrm{O7V}}$ is the Lyman continuum flux of an individual O7V
star.  
distance of 20 Mpc and using the flux values given in Table
\ref{tab:flux}, we obtain $\mathrm{L(H} \beta) = (5.0 \pm 1.8) \times
10^{39} \mathrm{erg\,s}^{-1}$ for knot~A and $\mathrm{L(H}\beta) =
(3.1 \pm 1.9) \times 10^{39}\mathrm{erg\,s}^{-1}$ for knot~B. Taking
$Q_0^{\mathrm{O7V}} = 10^{48.9}\,\mathrm{photons\,s}^{-1}$, as
suggested by \citet{hadfield06} for LMC metallicity O stars, implies
that knot~A contains $1300 \pm 500$ equivalent O7V stars and that
knot~B contains $800 \pm 500$ equivalent O7V stars.
 The Starburst99 modelling predicts an $\mathrm{H}\beta$ luminosity
 for cluster~1 of $(4.5 \pm 1.1) \times 10^{39} \mathrm{erg\,s}^{-1}$,
 showing that cluster~1 dominates knot~A in terms of its
 $\mathrm{H}\beta$ luminosity. This implies the presence of $1200 \pm
 300$ equivalent O7V stars within cluster~1.
  This number of equivalent O7V stars can be converted into the total
  number of O stars, $N(\mathrm{O})$, using the time-dependent
  parameter, $\eta(t)$:
$$ N(\mathrm{O})=\frac{N(\mathrm{O7V})}{\eta(t)}.
$$
For an age of 5 Myr and a metallicity of $Z=0.008$, \citet{schaerer98}
give $\eta = 0.2$, implying that cluster~1 contains $5900 \pm 1400$ O
stars. As knot~A is dominated by cluster~1, applying $\eta=0.2$
provides a good approximation to the number of O stars within knot~A
of $6600 \pm 2400$ O stars.  For knot~B, which is dominated by the
young cluster~5, the total number of O stars is comparable to the
number of equivalent O7V, since $\eta \sim 1$ for ages of $< 3$ Myr
\citep{schaerer98}.

\subsection{Wolf-Rayet star content}
\label{sec:WR}

\begin{figure}
\includegraphics[angle=270,width=9cm]{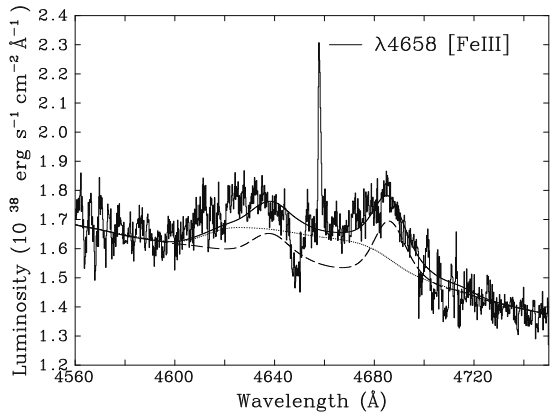}
\caption{\label{fig:wr} The $\lambda 4640$ Wolf-Rayet blue bump in the
spectrum of knot~A. The solid line shows the best empirical fit to the
spectrum, of 550 WNL stars and 200 WCE stars. The dashed line shows
only that from 550 WNL stars and the dotted line shows only the
contribution from 200 WCE stars. Nebular [Fe~III] $\lambda 4658$
emission is marked.}
\end{figure}

The number of WR stars within knot~A can be obtained by considering
the luminosity of the blue WR bump. The similarity in the strengths of
N~III 4640 and He~II 4686 in knot~A suggests the presence of late-type
WN (WNL) stars.  The luminosity of $\lambda 4686$ is $6.8 \times
10^{38} \mathrm{erg\,s}^{-1}$, after correcting for an extinction of
$E(B-V)_{\mathrm{int}} = 0.24$~mag and a distance of 20 Mpc. Comparing
this to the average luminosity of LMC WN7-9 stars of ($7.2\,\pm\,6.7)
\times 10^{35} \mathrm{erg\,s}^{-1}$ \citep{crowther06}, indicates
that knot~A contains 950 WNL stars, assuming no contribution to the
blue bump from WC stars. However, \citet*{guseva00} identify early-type 
WC (WCE) stars 
in their long-slit spectrum of the entire central region of NGC~1140,
by the presence of $\mathrm{C~IV}\,5808$ emission.
 Therefore, we empirically match the blue bump using various multiples
 of an LMC WN7-9 star, plus an LMC WC4 star, finding a best fit of
 approximately $550\,\pm\,100$ WNL stars and $200\,\pm\,50$ WCE stars.
 This is shown in Fig. \ref{fig:wr}.  We assume that all of the WR
 stars within knot~A are contained in cluster~1, since it is the most
 massive young cluster within this region.

\subsection{Star formation rate of NGC~1140}

We measure an observed $\mathrm{H}\alpha$ flux of $\mathrm{F(H}\alpha)
= (1.2 \pm 0.2) \times
10^{-12}\,\mathrm{erg\,s}^{-1}\,\mathrm{cm}^{-2}$ for the whole
galaxy, based upon aperture photometry of the continuum subtracted
F658N image (see Section \ref{sec:imaging}).
Assuming a distance of 20 Mpc and a mean extinction of $E(B -
V)_{\mathrm{int}} = 0.19$~mag, we find an $\mathrm{H}\alpha$
luminosity of NGC~1140 of $(9 \pm 3)\times
10^{40}\,\mathrm{erg\,s}^{-1}$. This mean extinction was determined as
an average of the nebular extinctions of knots~A and B (Section
\ref{sec:fluxes}) weighted by the relative contributions to the
$\mathrm{H}\beta$ flux of the central region of 2:1 (taken from Table
\ref{tab:flux}).
Using the equation:
\begin{eqnarray*}
\mathrm{SFR (M}_\odot \mathrm{yr}^{-1}) &=& 7.9 \times 10^{-42} \times
                       \mathrm{L(H}\alpha) \mathrm{(erg\,s}^{-1}) \\
                       &=& 1.08 \times 10^{-53} \times
                       Q_0^{\mathrm{Obs}} (\mathrm{photons\,s}^{-1})
\end{eqnarray*}
 \citep{kennicutt98}, we estimate the SFR of the
 galaxy of $0.7 \pm 0.3\,\mathrm{M}_\odot \mathrm{yr}^{-1}$, in good
 agreement with the SFR of $0.8\,\mathrm{M}_\odot\,\mathrm{yr}^{-1}$,
 obtained from ground-based continuum subtracted $\mathrm{H}\alpha$
 imaging by \citetalias{hunter94_vanwoerd}, after adjustment to our
 adopted distance.  Our $\mathrm{H}\alpha$ luminosity gives
 $Q_0^{\mathrm{Obs}} = ( 6.9 \pm 2.6) \times
 10^{52}\,\mathrm{photons\,s}^{-1}$ or $9000 \pm 3000$ O7V equivalents
 in NGC~1140.

\section{Dynamical masses of the clusters}
\label{sec:mass}

The virial motions of the individual stars within a cluster, as
measured by the line-of-sight velocity dispersion, give a dynamical
estimate of the cluster mass. This also requires a knowledge of the
half-light radius of the cluster, which is the two-dimensional radius
within which half of the projected luminosity of the cluster is
contained.

\subsection{Half-light radius}
\label{sec:rhl}

The half-light radii of clusters 1 and 6 were determined from the
 drizzled ACS/\emph{HST} F625W images using the program {\sc ishape}
 \citep{larsen99}.  The routine convolves the point-spread function
 (PSF) of the camera with the desired light profile.  {\sc ishape}
 computes a $\chi^2$ minimisation between the cluster image and the
 model profile over a set fitting radius, iterating the FWHM, the
 ratio of the minor to major axes and the orientation of the fit.  In
 order to obtain the best fit, the user specifies the `clean radius',
 which is the largest radius that contains no contamination by
 neighbouring clusters.

Cluster~6 is more isolated than cluster~1, and is uncontaminated by
 its neighbours up to a clean radius of four pixels (1 pixel = 0.05
 arcsec).  Cluster~1, however, is only isolated up to two pixels.  The
 half-light radii were computed by {\sc ishape} for a range of
 profiles, using a PSF that was computed in Tiny Tim \citep{krist97},
 for fitting radii between two and twenty pixels with a clean radius
 of two pixels for cluster~1 and four pixels for cluster~6.  The
 profiles considered were a Gaussian profile, EFF profiles
 \citep*{elson87} of index $x$, which take the form:
$$ f(z) = \frac{1}{(1+z^2)^{\gamma}}\,, \qquad \gamma = 0.1x
$$ and \citet{king62} profiles of index $c$ of the form:
$$ f(z) = \left \{
\begin{array}{ll} 
\left ( \frac{1} {\sqrt{1+z^2}}-\frac{1}{\sqrt{1+c^2}} \right )^2 & z
< c \\ 0 & z \ge c\,. \\
\end{array}
\right.
$$

\begin{table}
\caption{\label{tab:radius} Mean half-light radii computed by {\sc
ishape} for a range of model profiles for fitting radii between two
and eight pixels, and the corresponding standard deviations
(s.d.). Half-light radii were adopted on the basis of the best
standard deviation.}
\begin{center}
\begin{tabular}{lccccc} \hline

           & & \multicolumn{2}{c}{Cluster~1} &
           \multicolumn{2}{c}{Cluster~6}\\
Model & Index &$\mathrm{r}_{\mathrm{hl}}$ (pc)& s.d. &
$\mathrm{r}_{\mathrm{hl}}$ (pc)& s.d. \\ \hline
Gaussian & --- & 6.1 & 0.8 & 5.6 & 0.4 \\ EFF & 15 & 9.5 & 0.5 & 8.4 &
0.5 \\ EFF & 25 & 6.9 & 0.8 & 6.12 & 0.12 \\ King & 5 & 6.6 & 0.8 &
5.79 & 0.12 \\ King & 15 & 8.2 & 0.5 & 7.0 & 0.6 \\ King & 30 & 10.4 &
0.4 & 8.8 & 1.2 \\ King & 100 & 17.3 & 0.6 & 14.6 & 2.6 \\ \hline
Adopted & & \multicolumn{2}{l}{\hspace{0.2cm} $8\,\pm\,2$} &
\multicolumn{2}{l}{\hspace{0.15cm} $6.0\,\pm\,0.2$} \\ \hline
\end{tabular}
\end{center}
\end{table}

There was a sudden jump in the values of half-light radii produced for
fitting radii of eight pixels and for nine pixels. Therefore, the
means of the half-light radii computed by {\sc ishape} for each
profile for fitting radii between two and eight pixels were
considered. These, along with the corresponding standard
deviation, are contained in Table~\ref{tab:radius}. As a good profile
fit will not vary much with fitting radius, the profile with the
lowest standard deviation was adopted. For cluster~1 this was the
King~30 profile. However, the standard deviation of this is quite
large and similar to that produced for other profiles. Discarding the
outlying values produced by the King 100~profile, the mean and
standard deviation of the 42 results produced by the other 6 profiles
gives $\mathrm{r}_{\mathrm{hl}} = 8\,\pm\,2$ pc for cluster~1.

For cluster~6, both the EFF 25 and the King 5 profile show an equally
high level of consistency and a mean of the results of both of these
profiles was adopted to give $\mathrm{r}_{\mathrm{hl}} =
6.0\,\pm\,0.2$ pc for cluster~6.

\subsection{Velocity dispersion}
\label{sec:veldisp}

The line-of-sight velocity dispersion, $\sigma$, of a cluster can be
determined by comparing the broadening in the lines of the cluster
spectrum with respect to those in a red supergiant template
spectrum. There are two main methods that quantify the comparison, and
these are discussed below.

 The spectrum is compared to red supergiant templates because the
  lines of these stars are broadened by only a few
  $\mathrm{km\,s}^{-1}$, by macro-turbulence in their atmospheres
  \citep{gray86}. It is not appropriate to use earlier type
  supergiants and main sequence stars, as effects such as rotational
  broadening, macro-turbulence and micro-turbulence broaden the lines
  of these stars by amounts comparable to the anticipated cluster
  velocity dispersions.  Therefore, only spectral regions redwards of
  $\sim 5000 \textrm{\AA}$, which are dominated by light arising from
  cool supergiants of spectral type F-M, should be considered (see
  e.g. \citealt{ho96a}). Suitable spectral regions should show visible
  similarity between the cluster and RSG spectra and should also not
  contain any telluric lines, which provide an artificial match
  between the spectra.

 The first method relies on minimising a reduced $\chi^2$ between the
normalised cluster spectrum and normalised, broadened template
spectra. The broadening is achieved by convolving the normalised
template spectrum with a Gaussian of $\sigma$ equal to the desired
velocity broadening.  The broadened template is multiplied by an
optimum factor, so that it produces the lowest possible
reduced-$\chi^2$ between the broadened template spectrum and the
cluster spectrum.  The template is broadened by a range of suitable
values and a reduced-$\chi^2$ is similarly computed for each value of
broadening.  The lowest of these $\chi^2$ values determines what
amount of broadening produces the best match with the cluster for a
given template, and so indicates the velocity dispersion of the
cluster. This is repeated for a range of template spectral types. This
method relies on a good match between the relative line strengths of
the cluster and template and can thus be sensitive to the spectral
type of the template star.

The second method utilises the cross-correlation technique of
  \citet{tonry79}. It requires that the spectra being considered are
  normalised, and continuum subtracted, to give a flat continuum at
  zero.  The cluster is then cross-correlated with a red supergiant
  template over suitable spectral regions, and the FWHM of the
  resulting cross correlation function (CCF) is measured.  The
  template spectrum is broadened by a range of velocities. Each
  broadened template is cross-correlated with the original,
  unbroadened template and the FWHM of each CCF is measured. In this
  manner, the near-linear relationship between broadening and the FWHM
  of the CCF can be empirically calibrated to an absolute scale. This
  calibration is applied to the FWHM of the original CCF, produced by
  cross-correlating the cluster spectrum with the template spectrum,
  to determine the velocity dispersion of the cluster.  This is
  repeated for each template star. The conversion factor differs with
  template spectrum.  While this method is less sensitive to spectral
  type matching, it suffers from complications associated with the
  subjectivity of fitting CCFs. These include factors such as
  selecting the background level and fitting non-Gaussian CCFs, and
  are especially important when the CCF is weak.

\begin{table}
\caption{\label{tab:veldisp} Values of line-of sight velocity
dispersion over the region $8485-8845 \textrm{\AA}$ for each red
supergiant template observed. The templates are identified by their
\citet{massey03} catalogue number. It is marked whether the
cross-correlation (Xcor) or $\chi^2$ minimisation ($\chi^2$) technique
was used. The Paschen emission was masked out for the reduced $\chi^2$
minimisation technique. The results obtained for the cross-correlation
technique are an upper limit on the cluster velocity dispersion.}
\begin{center}
\begin{tabular}{llcccc} \hline
\multicolumn{2}{c}{RSG Template} & \multicolumn{4}{c}{Velocity
dispersion ($\mathrm{km\,s}^{-1}$)} \\ Cat No. & Spectral & Cluster1
&& \multicolumn{2}{c}{cluster~6} \\ & Type & $\chi^2$ && Xcor &
$\chi^2$ \\ \hline
64448 & K2-7 I & 23 && 26 & 26 \\ 
30840 & K3-5 I & 23 && 29 & 27 \\
66694 & K5 I & 23 && 30 & 26 \\ 
71566 & K7 I & 22 && 29 & 26 \\
20133 & M0 I & 24 && 31 & 27 \\ 
50840 & M1-2 I & 22 && 32 & 26 \\ \hline 
Mean & & 24 && 30 & 26 \\ \hline
\end{tabular}
\end{center}
\end{table}

Since clusters 1 and 6 are the brightest cluster members in knots A
and B, we assume that the velocity dispersions of these clusters
dominate the broadening of the RSG features apparent in the knots.

Unfortunately, the youth of cluster 1 means that the RSG features in
 the spectrum are very weak, and many of the lines visible in the
 template spectra are absent in the cluster spectrum.  The low
 signal-to-noise ratio of knot~B causes a similar problem.  The
 strongest RSG features, the Ca~II triplet absorption lines, are
 clearly visible in the cluster spectra.  However, these lines are
 saturated in the template spectra. As the core of a saturated profile
 is narrower than would have been produced in a Gaussian profile, and
 it is the core that produces the CCF signal, cross-correlation of
 these regions tends to overestimate the cluster velocity dispersion
 (see e.g. \citealt{walcher05}).  However, cross-correlation of all
 other regions that both contain RSG features and lack telluric
 features produce very noisy, non-Gaussian CCFs, which cannot be
 robustly fitted. Therefore, only the region $8485 - 8845
 \textrm{\AA}$ was considered for cross-correlation. Cross-correlation
 of even this region of knot~A with the templates did not produce CCFs
 that could be confidently fitted, likely due to the contamination of
 the Ca~II triplet by the strong Paschen lines. Therefore, no results
 were produced for knot~A using the cross-correlation technique.

The reduced-$\chi^2$ minimisation could also only be computed for the
region $8485 - 8845 \textrm{\AA}$, due to the lack of strong RSG
absorption lines except around the Ca~II triplet. The Paschen emission
features were masked out of both knot spectra before the
reduced-$\chi^2$ minimisation was computed. The results of the
cross-correlation and the reduced-$\chi^2$ minimisation are listed in
Table~\ref{tab:veldisp}.

The results show consistency over all the spectral types, with the
cross-correlation of cluster~6 with the templates producing
systematically higher velocity dispersions than the reduced-$\chi^2$
technique, as expected. There should not be any systematic
uncertainties in the velocity dispersion calculated by
reduced-$\chi^2$ minimisation, despite the use of the Ca~II triplet
lines. \citet{mengel02} found no disparity between the velocity
dispersion results computed by $\chi^2$ minimisation for the strongest
component of the Ca~II triplet and other individual absorption
features for clusters in NGC~4038/4039. The mean and standard
deviation of the velocity dispersions calculated from the
reduced-$\chi^2$ minimisation from all six template stars give values
of $\sigma = 24\,\pm\,1\,\mathrm{km\,s}^{-1}$ for cluster~1 and
$\sigma = 26\,\pm\,1\,\mathrm{km\,s}^{-1}$ for cluster~6.

\subsection{Virial mass}

The virial equation relates the virial mass of a cluster, $M_{\mathrm{dyn}}$, 
to the
line-of-sight velocity dispersion, $\sigma$, 
and the half-light radius, $r_{\mathrm{hl}}$, of the cluster by the
equation:
$$ M_{\mathrm{dyn}} \approx \frac {\eta \sigma^2 r_{\mathrm{hl}}}{G}\,.
\label{eqn:M}
$$
 This equation assumes that a cluster is gravitationally bound,
spherically symmetrical and virialised, that the velocity dispersion
of the cluster is isotropic and that all of the stars contained in the
cluster are single stars of equal mass.  \citet{spitzer87} showed that
$\eta=9.75$ for globular clusters with a wide range of light profiles.
 However, mass segregation and the presence of binary stars within the
 cluster can cause a large variation in the value of $\eta$
(\citealt{boily05,fleck06}; Kouwenhoven \& de Grijs, in prep.).
 For the young age and high
 mass of cluster 1, however, these effects are minimal.
Table~\ref{tab:clusprop} summarises the cluster properties.

\begin{table}
\caption{\label{tab:clusprop} Summary of cluster properties of
clusters 1 and 6.}
\begin{tabular}{lr@{$\pm$}lr@{$\pm$}l} \hline
 Property &\multicolumn{2}{c}{cluster~1} & \multicolumn{2}{c}{cluster~6} \\ \hline 
$E(B-V)$(cluster) (mag)         & 0.24 & 0.04  & \multicolumn{2}{c}{---}\\ 
Age (Myr)                       & 5    & 1     &\multicolumn{2}{c}{---}\\ 
$\mathrm{r}_{\mathrm{hl}}$ (pc) & 8    & 2     &6.0 & 0.2 \\ 
$\sigma\,(\mathrm{km\,s}^{-1})$ & 24   & 1     & 26 & 1 \\
$M_{\mathrm{dyn}} (10^6\,\mathrm{M}_\odot$) 
                                & 10   & 3     & 9.1 & 0.8 \\
$L_V (10^7\,\mathrm{L}_\odot)$  & 7.3  &1.8    & \multicolumn{2}{c}{---} \\
$L_V/M_{\mathrm{dyn}} ((L_V/M)_\odot)$ 
                                & 7.0 & 2.6    & \multicolumn{2}{c}{---} \\ \hline
\end{tabular}
\end{table}

\section{Discussion}

 In this section we discuss the disparity found between the dynamical
 and photometric mass estimates. We also consider the reliability of
 stellar evolution models by comparing them with observations.

\subsection{How reliable are photometric mass estimates versus dynamical mass estimates?}

\begin{figure}
\includegraphics[width=8.5cm]{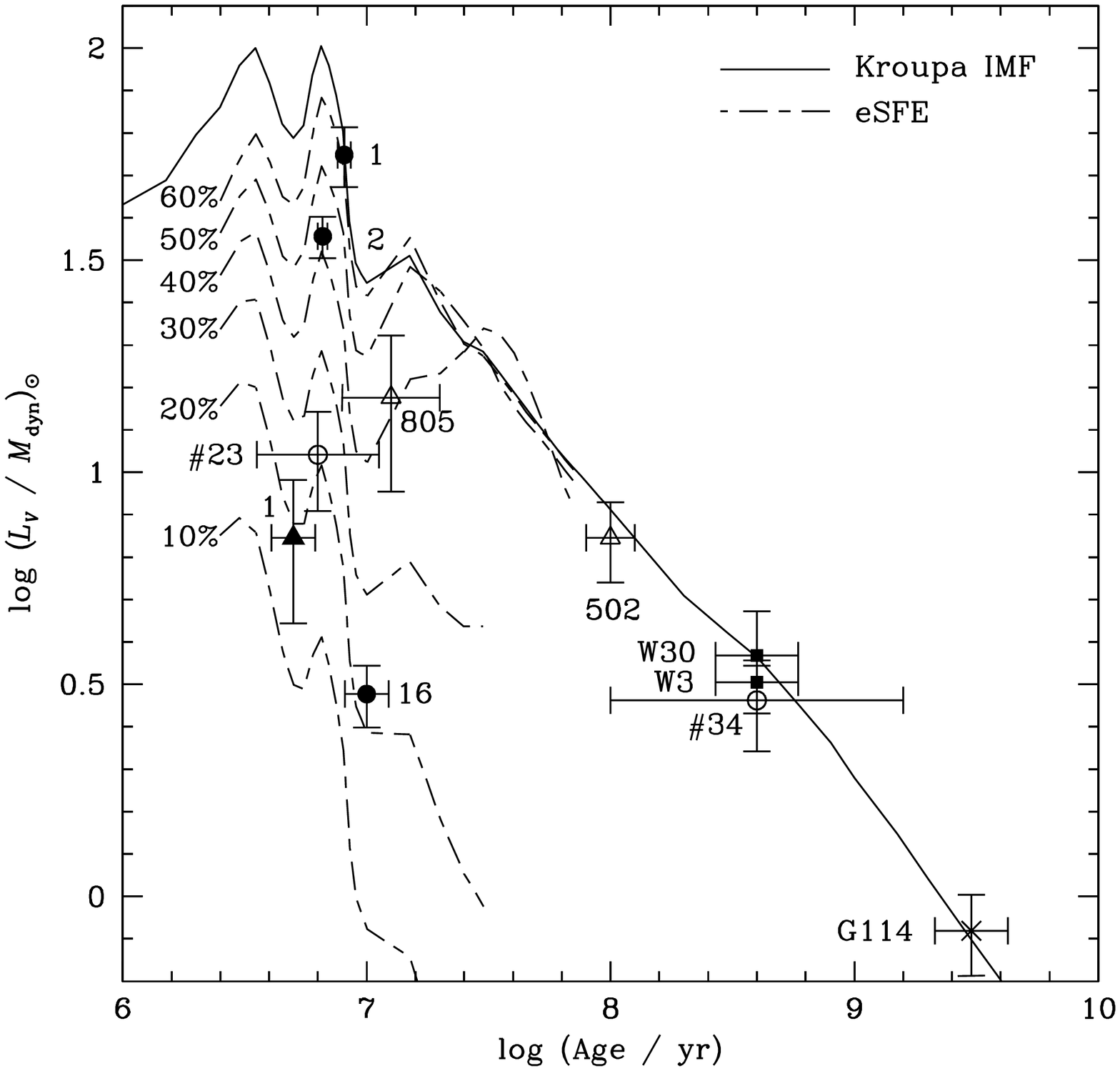}
\caption{\label{fig:lm} Diagram of light to dynamical mass ratio
against age for cluster~1 in NGC~1140 (filled triangle; this work),
and for clusters 1, 2 and 16 in NGC 4038/4039 (filled circles;
\citealt{mengel02}), clusters 502 and 805 in NGC 5236 (open triangles;
\citealt{larsen04_rich}), clusters W3 \citep{maraston04} and W30
\citep{bastian06_sag} in NGC 7252 (filled squares), cluster G114 in
NGC 1316 (cross; \citealt{bastian06_sag}) and clusters $\#23$ and
$\#34$ in ESO 338-IG04 (open circles; \citealt{ostlin06}).  Also plotted
are the model evolution of light-to-mass ratio predicted by a Kroupa
IMF with \citet{maraston05} SSPs for a solar metallicity burst and the
effect that non-$100\%$ effective star-formation efficiencies (eSFEs)
have on the model light-to-mass ratio \citep{goodwin06_bast}.}
\end{figure}

 The dynamical mass determined for cluster~1 is several times greater
than the photometric mass found.  To assess this disparity,
Fig. \ref{fig:lm} compares the light-to-dynamical mass ratio of the
cluster to a model light-to-mass ratio at the known age of the
cluster. The model assumes \citet{maraston05} SSPs, a Kroupa IMF and
solar metallicity. Data for several other clusters whose velocity
dispersions were measured from UVES observations are also
included. 
It is clear from this figure that many other young ($< 20 $ Myr)
clusters also have light-to-dynamical mass ratios well below model
predictions, while the older ($> 20 $ Myr) clusters have ratios that
agree well with the canonical value.

As discussed by \citet{bastian06_sag} and \citet{goodwin06_bast}, the
 most likely explanation for the discrepancy between model and
 observations seen in the sample of young clusters is a lack of virial
 equilibrium caused by violent relaxation after the formation of the
 cluster and from the expulsion of gas. This causes the measured
 dynamical mass to overestimate the true mass.  Since the older
 clusters, which are expected to be in virial equilibrium after
 surviving gas expulsion, lie on or near the model line, the
 photometric mass is likely to be a good representation of the true
 cluster mass. This supports the assumption of a standard IMF.  In
 this case, the disagreement between the dynamical mass and the
 photometric mass of a cluster can be used to assess to what degree
 the cluster is out of virial equilibrium.  This can be parameterised
 by the effective star formation efficiency (eSFE) of the cluster,
 $\epsilon$; at the onset of gas expulsion, a cluster with an
 $\mathrm{eSFE} = \epsilon$ has a velocity dispersion that is
 $\sqrt{1/\epsilon}$ too large to be in virial equilibrium
 \citep{goodwin06_bast}. The model light-to-mass ratio assuming a
 range of effective star formation efficiencies are included in
 Fig. \ref{fig:lm}, taking the onset of gas expulsion as 2 Myr. These
 tracks differ from those presented by \citet{goodwin06_bast} as they
 are not smoothed.

An alternative explanation for clusters lying below the canonical line
in Fig.  \ref{fig:lm} is that these clusters have non-standard IMFs.
In this situation, the dynamical mass represents the true cluster
mass, and the incorrect assumption of a standard IMF causes the
photometric mass to differ from the dynamical mass.  Since the
discrepant clusters lie below the model line, they would have an
over-abundance of low-mass stars with respect to the Kroupa IMF. Such
bottom-heavy clusters are more likely to remain bound than clusters
with a standard Kroupa IMF and should be apparent in the sample of old
clusters. As they are not, this explanation seems unlikely for the
majority of these young clusters. However, a non-standard IMF can not
be ruled out in the case of NGC~1140--1.

There are various potential explanations for the difference between
light-to-dynamical mass ratio and model light-to-mass ratio in the
case of NGC~1140--1. These are explored below:
\begin{enumerate}
\item{Assuming that the difference bewteen dynamical and photmetric
mass is due solely to the cluster being out of virial equilibrium, it
can be seen from Fig. \ref{fig:lm} that cluster~1 has an eSFE of
around $10-20 \%$. \citet{goodwin06_bast} predict that a cluster with
an eSFE of $10 \%$ will lose $80\%$ of its mass in $\sim 20$ Myr and
not remain bound. As such, the cluster would disperse over a
relatively short timeframe and, therefore, not evolve into a second
generation globular cluster.}
\item{An underestimate of the cluster extinction would have caused an
underestimation of the photometric mass.  However, in order to
increase the luminosity of the cluster by the factor of $\sim 9$
required to match the photometric mass with the dynamical mass, the
extinction would need to have been underestimated by around
$E(B-V)_{\mathrm{int}} \approx 0.9$~mag.  Since the extinction
calculated is in good agreement with previous measurements, and is
already larger than the nebular extinction derived, this is unlikely
to be the case.}
\item{Determining the half-light radii of the clusters was heavily
dependent on the light profile adopted and several of the profiles
were in good agreement with observation. However, adopting
$\mathrm{r}_{\mathrm{hl}} \approx 6.0$ pc, the lowest potential
half-light radius computed by {\sc ishape} for cluster~1, would still
produce a virial mass that was a factor of seven larger than the
photometric mass.}
\item{The crowded nature of the region may have caused the cluster
 velocity dispersion to be overestimated.  The broadening of the RSG
 features seen in the knot spectra may not have been due solely to the
 virial motions of the brightest cluster within the knot, as is
 assumed, but may have had contributions from the motions of other
 clusters. This situation can be modelled using the template spectra,
 and a reduced $\chi^2$ minimisation can be computed for this model
 knot.  A template spectrum is broadened by $8\,\mathrm{km\,s}^{-1}$
 to represent the broadening of cluster~1, as implied by its
 photometric mass and half-light radius. Added to this is the
 cluster~2 contribution, in the flux ratio of 1/1.7, as implied by the
 F814W photometry in Table~\ref{tab:mag}. Taking the velocity
 dispersion of cluster~2 to be equal to that of cluster~1, the
 velocity dispersion found for knot~A in Table~\ref{tab:veldisp} can
 be reproduced by introducing a relative velocity shift between the
 two clusters of $\sim 20\,\mathrm{km\,s}^{-1}$.  It is not possible
 to assess how likely such a shift is.  Cluster 2 appears to be older
 than cluster~1 because of the lack of nebular emission
 (Fig. \ref{fig:obs}). This then implies that cluster~2 could have a
 larger photometric mass than cluster~1, despite its fainter
 luminosity, and thus have a larger velocity dispersion than adopted
 here. This would reduce the velocity shift required to produce a
 combined velocity dispersion equivalent to that observed for knot A.
 However, neither the photometric mass nor the shift are known and the
 importance of the contribution of cluster 2 to the velocity
 dispersion of knot~A cannot be assessed.}
\end{enumerate}
It is not possible to quantify the degree to which these four factors
contribute to the difference between the light-to-dynamical mass ratio
measured and the model in Fig. \ref{fig:lm}. However, the contribution
from the neighbours of cluster~1 is likely to be the most dominant
factor for the large dynamical mass measured. As it is not possible to
quantify this, it is also not possible to assess how long the cluster
is likely to survive.

\subsection{Young clusters as testing grounds of stellar evolution models}

\begin{table*}
\caption{\label{tab:knotprop} Nebular knot properties and massive star
content of NGC~1140 (this work), NGC~3125 \protect \citep{hadfield06}
and Tol~89 \citep{sidoli06}. }
\begin{tabular}{lr@{$\pm$}lr@{$\pm$}lccccccc} \hline
Galaxy & \multicolumn{4}{c}{NGC~1140}                    && \multicolumn{3}{c}{NGC~3125}   && \multicolumn{2}{c}{Tol 89 (NGC~5398)} \\ \hline Knot
       &\multicolumn{2}{c}{A}    &\multicolumn{2}{c}{B}  && \multicolumn{2}{c}{A}    & B   && A    & B \\ \hline
$E(B-V)$ (mag)                   & 0.16&0.06 & 0.23&0.08 && \multicolumn{2}{c}{0.16} &0.13 && 0.07 & 0.29\\ 
$N_e\,\mathrm{(cm}^{-3})$        & 60&50      &\multicolumn{2}{c}{---}&& \multicolumn{2}{c}{140}  & 140  &&$90\pm40$& $150\pm40$ \\ 
$T_e$ (K)                        & 9700&300   &\multicolumn{2}{c}{---}&& \multicolumn{2}{c}{10500}& 9800 && $10000\pm300$ & $9800\pm300$ \\
$12+\log{\mathrm{O/H}}$          & 8.29&0.09  &\multicolumn{2}{c}{---}&& \multicolumn{2}{c}{8.32} & 8.35 &&$8.27^{+0.08}_{-0.09}$&
$8.38^{+0.06}_{-0.07}$ \\ $N\mathrm{(O)}^\star$ & 6600&2400 & 800&500 && \multicolumn{2}{c}{4000} & 3200 && 685 & 2780 \\ 
Age (Myr) & 5&1                  &\multicolumn{2}{c}{$>10$}           && \multicolumn{2}{c}{$\sim 4$} &$\sim 4$ && 3.5-4 & $< 2.5$ \\ \hline 
Cluster &\multicolumn{2}{c}{1}   &\multicolumn{2}{c}{6} && A1 &A2 & B1 + B2 && A1 & B1 \\
\hline $M_{\mathrm{phot}} (10^6 \mathrm{M}_\odot$) & 1.1&0.3 &\multicolumn{2}{c}{---} && 0.20 & 0.22 & 0.16 && $0.1-0.2$ & $\sim 0.03$ \\ 
$N(\mathrm{O)}^{\dagger}$        & 5900&1400 &\multicolumn{2}{c}{---} && 550 & 600 & 450 && $310-680$ & $\sim 100$ \\ 
$N\mathrm{(WN)}$                 &550&100 &\multicolumn{2}{c}{---} && 105 & $\sim 55$ & 40 && 80 & ---\\ 
$N\mathrm{(WC)}$                 &200& 50 &\multicolumn{2}{c}{---} && 20 & --- & 20 && 0 & --- \\ 
$N\mathrm{(WR)}/N\mathrm{(O)}$   &\multicolumn{2}{c}{0.1}   &\multicolumn{2}{c}{---} && 0.2 & 0.1 & 0.1 && 0.2 & --- \\ \hline
\multicolumn{12}{l}{$^\star$ From the luminosity of the nebular $\mathrm{H} \beta$ emission of the knot.} \\
\multicolumn{12}{l}{$^{\dagger}$ From Starburst99, for the age and photometric mass of the cluster.} \\
\end{tabular}
\end{table*}

The nebular properties of the two knots of NGC~1140, and the massive
star content of clusters 1 and 6 are presented in Table
\ref{tab:knotprop}.  There have been several other studies of
young massive stellar populations of nearby starbursts, including the
ultraviolet (UV) survey of \citet*{chandar04} and optical studies of \citet{vacca92}
and \citetalias{schaerer99a}. 
\defcitealias{schaerer99a}{Schaerer et al. (1999a)}
 However, the results of these surveys are not readily comparable
 to our results, due to the different techniques adopted, especially
 in determining the WR populations.

\citet{chandar04} derived the O and WR content of the central
starburst regions for a large sample of galaxies based upon UV
\emph{HST} spectroscopy and Starburst99 models. Fits to far-UV
spectral morphologies and continuum slopes provided representative
ages and extinctions from which O star populations were obtained,
although individual clusters were not distinguished. The number of WR
stars was inferred from a calibration of He~II 1640 line
luminosities.  Individual clusters were also not considered in the
ground-based optical studies of \citet{vacca92} and
\citetalias{schaerer99a}. Standard nebular techniques were applied to
derive extinctions, representative ages and O star numbers. WR
populations resulted from calibrations of Galactic and LMC WR stars
(\citealt{vacca94b,schaerer98}), which  differ from our calibration with 
 solely LMC templates.

Studies of individual massive clusters using common techniques,
including \emph{HST} imaging, have been carried out by
\citet{hadfield06} and \citet{sidoli06} for two galaxies of LMC-like
 metallicity, the blue compact dwarf galaxy NGC~3125 and the giant HII
region (GHR) Tol~89 located in the barred spiral galaxy NGC~5398. A
comparison of individual knots, as derived from nebular properties,
and cluster masses and O star numbers, from Starburst99 modelling, is
made in Table~\ref{tab:knotprop}. WR populations follow from optical
calibrations of \cite{crowther06} in all cases.

 NGC~1140--1 is a  significantly more massive counterpart to
NGC~3125--A1 and Tol~89--A1, with a similar, high ratio of WR to
O type stars, $N(\mathrm{WR})/N(\mathrm{O}) \sim 0.1-0.2$ and
$N(\mathrm{WC})/N(\mathrm{WN}) \le 0.4$.  These empirical
stellar populations can be compared to predictions from evolutionary
synthesis models (e.g. Starburst99), for which reduced WR populations
of $N(\mathrm{WR})/N(\mathrm{O}) \sim 0.02 - 0.1$ and a substantially
different subtype distribution of $N(\mathrm{WC})/N(\mathrm{WN}) \sim
40$ are predicted.  These are based on standard Geneva evolution
models \citep{meynet94} for ages of $4-5$~Myr (see also
\citealt{sidoli06}).  Allowance for rotational mixing in evolutionary
models does not resolve these discrepancies,
i.e. $N(\mathrm{WR})/N(\mathrm{O}) < 0.1$ and
$N(\mathrm{WC})/N(\mathrm{WN}) \sim 10$ according to \citet{vazquez07}
for a 5 Myr old population at $Z~=~0.008$.  Padova models
\citep{fagotto94} also predict a low ratio of WR to O stars, with
$N(\mathrm{WR})/N(\mathrm{O}) \sim 0.04 - 0.07$.  These models,
however, predict significantly more realistic ratios of WC to WN stars
of $N(\mathrm{WC})/N(\mathrm{WN}) \sim 0.2-2$ between $4-5$~Myr.

\section{Summary}

We present new high spectral resolution VLT/UVES spectroscopy and
 \emph{HST}/ACS imaging of the central region of NGC~1140. It is
 apparent from the ACS imaging that the central region contains
 several clusters, although this is only resolved into two
 star-forming knots by the UVES spectroscopy: knot~A, which contains
 clusters 1 and 2, and knot~B, which contains clusters 5, 6 and 7.

Nebular analysis of knot~A yields an LMC-like metallicity of $12 +
\log\mathrm{O/H} = 8.29\,\pm\,0.09$.  Starburst99 modelling indicates an age of
$5\,\pm\,1$ Myr and a photometric mass of $(1.1\,\pm\,0.3) \times
10^6\,\mathrm{M}_\odot$ for cluster~1. Virial masses of $(10 \pm 3)
\times 10^6\,\mathrm{M}_\odot$ and $(9.1 \pm 0.8) \times
10^6\,\mathrm{M}_\odot$ were determined for clusters 1 and 6,
respectively, using the half-light radii determined from the F625W ACS
image and the velocity dispersions determined by computing a
reduced-$\chi^2$ minimisation between the cluster spectrum and RSG
template spectra over a spectral region containing the Ca~II
triplet. We interpret the difference between the dynamical and
photometric mass of cluster 1 as due to the crowded nature of knot A:
the velocity dispersion measured may not relate only to cluster 1, as
assumed, but likely contains a component that arises from cluster 2,
with a different systemic velocity to cluster 1.

We find 6600 and 800 O stars within knots~A and B, respectively, from
the $\mathrm{H} \beta$ luminosities of the knots. Our Starburst99
model predicts 5900 O stars within cluster~1. This implies that
$>90\%$ of the O stars within knot A are contained in cluster
1. Empricial fitting of the blue bump of cluster~1 indicates that this
cluster contains around 550 WN stars and 200 WC stars, giving ratios
of $N(\mathrm{WR})/N(\mathrm{O}) = 0.1$, if all of the WR stars lie
within cluster~1, and $N(\mathrm{WC})/N(\mathrm{WN}) = 0.4$. The
observed ratio of WR stars to O stars is significantly larger than
predicted by current evolutionary models. The observed ratio of WC to
WN stars is reproduced more successfully using Padova instantaneous
burst models than Geneva models, even allowing for rotational mixing.

\section*{Acknowledgments}

SM acknowledges financial support from PPARC/STFC.
The Image Reduction and Analysis Facility {\sc iraf} is distributed by
the National Optical Astronomy Observatories, which is operated by the
Association of Universities for Research in Astronomy, Inc., under
cooperative agreement with the U.S. National Science Foundation.
We would like to thank S{\o}ren Larsen for his help with the {\sc
ishape} analysis and Fabrizio Sidoli, who provided help with the UVES
data reduction. We would also like to thank Nate Bastian and Simon
Goodwin for supplying their effective star formation efficency models
and for discussions on the subject.

\bibliographystyle{mn2e} \bibliography{abbrev,refs}

\label{lastpage}

\end{document}